\documentclass[12pt]{article}
\usepackage{amsmath, amssymb, slashed, epsf, color, graphicx}

\begin{document}
\begin{titlepage}
\begin{center}

\hfill IPMU-14-0047 \\

\vspace{2.0cm}
{\large\bf Gaugino coannihilations}

\vspace{2.0cm}
{\bf Keisuke Harigaya},
{\bf Kunio Kaneta},
and
{\bf Shigeki Matsumoto}

\vspace{1.5cm}
{\it Kavli IPMU (WPI), University of Tokyo, Kashiwa, Chiba 277-8583, Japan}

\vspace{2.5cm}
\abstract{
The high-scale supersymmetry (SUSY) breaking scenario is now attracting many attentions, because it is consistent with almost all experiments of particle physics, astrophysics, and cosmology performed so far: e.g. it is possible to explain the Higgs mass of about 126\,GeV and contains WIMP dark matter candidates. In the scenario, gauginos are predicted to be around the TeV scale, and thus within a kinematically accessible range of near future experiments. Calculation of the thermal relic abundance for gaugino (bino or wino) dark matter is then of particular importance in order to clarify its mass consistent with cosmology and to determine future directions for exploring the high-scale SUSY breaking scenario. In this article, we calculate the abundance of the gaugino dark matter, with especially focusing on various coannihilations between gauginos, which has not been extensively studied so far. Our calculation involves the Sommerfeld effect on wino and gluino annihilations, which is known to give significant contributions to their cross sections. Based on obtained results, we discuss some implications to gaugino searches at collider and indirect detection experiments of dark matter.}

\end{center}
\end{titlepage}
\setcounter{footnote}{0}

\section{Introduction}
\label{sec: introduction}

The high-scale supersymmetry (SUSY) breaking scenario is at present attracting many attentions because of the discovery of the standard-model-like Higgs boson~\cite{higgs discovery} and null-observations of new physics signals at the Large Hadron Collider (LHC) experiment. This scenario is defined as the one having a split-type spectrum: all scalar particles except the lightest Higgs boson have their masses of ${\cal O}(10$-$100)$\,TeV, while gauginos and/or Higgsinos still remain at the scale of ${\cal O}(0.1$-$1)$\,TeV. Such a spectrum is realized by simple supergravity mediation of SUSY breaking, where the scalar particles (and the gravitino) acquire their masses via tree level interactions, while gaugino masses are dominated by one-loop anomaly mediated contributions~\cite{AMSB}.
The origin of the Higgsino mass, so-called the $\mu$-term, is model-dependent. In some models, it is generated via a tree level interaction to an order parameter of $R$-symmetry breaking and becomes of the order of the gravitino mass~\cite{mu-term}. We focus on such models in following discussion.

The split spectrum has also several phenomenological advantages. First, it is compatible with the Higgs mass of about 126\,GeV~\cite{higgs mass}. Second, it is not only compatible with null-observations of new physics signals at the LHC, but also ameliorates the problem of too large SUSY contributions to flavor-changing neutral currents. Third, it is free from the gravitino problem~\cite{gravitino problem}, so that it is compatible with the successful leptogenesis scenario~\cite{leptogenesis}. Furthermore, in a class of models, even the Polonyi problem~\cite{polonyi problem} is evaded. Finally, it involves several candidates for dark matter. Because of the advantages, many concrete models have been proposed so far, e.g. the PeV-scale SUSY model~\cite{Wells:2004di}, the pure gravity mediation model~\cite{PGM}, the spread SUSY model~\cite{Hall:2011jd}, the minimal split SUSY model~\cite{minimal split}, etc~\cite{other splits}.

Among several sparticles in the high-scale SUSY breaking scenario, gauginos are clearly remarkable ones, for their masses are at most at ${\cal O}(1)$\,TeV and are within kinematically accessible ranges of (near) future experiments. Calculation of the thermal relic abundance for the gaugino (bino or wino) dark matter is then of particular importance. It gives an upper limit on its mass and eventually enables us to find future directions to explore the high-scale SUSY breaking scenario. The scenario predicts three types of dark matter regions: bino dark matter, wino dark matter, and gaugino coannihilation regions. Bino dark matter region has already been ruled out, since its thermal relic abundance is too high to be consistent with the observed one~\cite{Ade:2013zuv}. Wino dark matter region has already been studied well and it has been found that the wino mass should be less than 3.1\,TeV~\cite{Hisano:2006nn}. Gaugino coannihilation regions, where either bino or wino can be dark matter, are remaining ones we should evaluate. Precise calculations of the abundance in the regions is in fact mandatory for future directions of collider and dark matter experiments, because it gives not only an upper limit on the dark matter mass but also detailed information about masses of coannihilating particles such as the gluino.

In this article, we calculate the thermal relic abundance of the dark matter in the framework of the high-scale SUSY breaking scenario. We particularly focus on the gaugino coannihilation regions. Our calculation involves the Sommerfeld effect~\cite{sommerfeld effect} on wino and gluino annihilations, which is known to give significant contributions to their cross sections when their masses are heavier than about 1\,TeV. In next section (section~\ref{sec: gauginos}), we first discuss gaugino masses in the high-scale SUSY breaking scenario in some details, where we consider several possible contributions to the masses~\cite{vector matters}--\cite{Harigaya:2013asa}: anomaly mediated contributions, Higgsino threshold corrections, contributions from a vector-like matter, and those from a Peccei-Quinn (PQ) sector. We then definitely show that the coannihilation regions are indeed realized by various realistic setups of the high-scale SUSY breaking scenario. Detailed calculations of the abundance in the regions are presented in section~\ref{sec: abundance}, where we finally clarify bino, wino, and gluino masses consistent with current cosmological observations and collider experiments. Section~\ref{sec: summary} is devoted to the summary of our discussion.

\section{Gaugino masses}
\label{sec: gauginos}

In this section, we review in some details how gauginos of the minimal supersymmetric standard model (MSSM) acquire their masses in the framework of the high-scale SUSY breaking scenario. Couplings between the SUSY breaking field and MSSM gauge multiplets are suppressed, because the breaking field is assumed to be charged under some symmetry or a composite one in this scenario. Such an assumption has a great advantage on its cosmology, for we do not have infamous Polonyi problem~\cite{polonyi problem}. Gaugino masses are therefore absent at tree level, while they are generated by quantum corrections. In what follows, we first discuss how the gauginos acquire their masses within the framework of the MSSM: anomaly mediated contributions and Higgsino threshold corrections. We next discuss contributions to the gaugino masses in some extensions of the MSSM: those from a vector-like matter field and a PQ sector. With well-motivated extensions of the MSSM, it then turns out that gaugino masses should be treated as free parameters for gaugino phenomenology, which is nothing but the topic we will develop in section~\ref{sec: abundance}.

\subsection{Anomaly mediated contributions}

Even if the SUSY breaking field does not couple to the gauge multiplets, quantum corrections unavoidably generate gaugino masses, which is known as anomaly mediated contributions~\cite{AMSB}. These are evaluated as $M_\lambda^{\rm (AM)} = \beta(g^2) \, m_{3/2}/(2g^2)$ with $g$, $\beta(g^2)$, and $m_{3/2}$ being the gauge coupling constant, the beta function of $g^2$, and the gravitino mass, respectively. At one loop level in the MSSM, bino, wino, and gluino masses ($M_1^{\rm (AM)}$, $M_2^{\rm (AM)}$, and $M_3^{\rm (AM)}$) are explicitly given by
\begin{eqnarray}
M_1^{\rm (AM)} = \frac{g_1^2}{16\pi^2} \frac{33}{5} m_{3/2},
\qquad
M_2^{\rm (AM)} = \frac{g_2^2}{16\pi^2} m_{3/2},
\qquad
M_3^{\rm (AM)} = -\frac{g_3^2}{16\pi^2} 3 \, m_{3/2}.
\label{eq: AM}
\end{eqnarray}

The anomaly mediated contributions can be understood as a consequence of quantum anomaly related to the scale invariance~\cite{AMSB}: Let us formulate the supergravity theory by the superconformal formalism~\cite{superconformal}. In this formalism, the Weyl compensator with a non-zero Weyl weight is introduced, and then the action has the local superconformal symmetry. This local superconformal symmetry is reduced to the local supersymmetry by a gauge fixing. As a result, the Weyl compensator obtains an F-term which is as large as the gravitino mass. This non-zero F-term induces the gaugino masses. Couplings between the Weyl compensator and the gauge multiplets which contribute to the gaugino masses, the dependence of gauge kinetic functions on the Weyl compensator in other words, are absent at tree level, as can be easily understood by vanishing Weyl weights of gauge kinetic functions. At loop level, however, gauge kinetic functions have non-zero Weyl weights due to the anomaly of the scale invariance, namely running of gauge coupling constants. The gauge kinetic functions hence depend on the Weyl compensator~\cite{Kaplunovsky:1994fg}. Resulting gaugino masses are therefore proportional to their beta functions of gauge coupling constants and are eventually given by equation~(\ref{eq: AM}).

\subsection{Higgsino threshold corrections}

\begin{figure}[t]
\begin{center}
\includegraphics[width=0.5\linewidth]{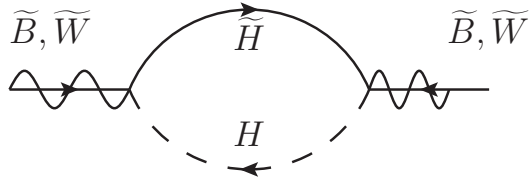}
\caption{\sl \small Higgsino threshold corrections to masses of electroweak gauginos.}
\label{fig: higgsinoth}
\end{center}
\end{figure}

Electroweak gauginos (bino and wino) receive threshold corrections from the Higgsino via the diagram in figure~\ref{fig: higgsinoth}. The corrections are evaluated as~\cite{AMSB},
\begin{eqnarray}
\Delta M_1^{\rm (HT)} = \frac{g_1^2}{16\pi^2} \frac{3}{5} L,
\qquad
\Delta M_2^{\rm (HT)} = \frac{g_2^2}{16\pi^2} L,
\qquad
L \equiv \frac{\mu \, m_A^2 \, \sin 2\beta}{|\mu|^2 - m_A^2}
\ln \frac{|\mu|^2}{m_A^2},
\label{eq: HT}
\end{eqnarray}
where $\mu$, $\tan\beta$, and $m_A$ are the Higgsino mass, the ratio of vacuum expectation values between up- and down-type Higgs doublets, and the mass of the heavy Higgs boson, respectively. The contributions become comparable to those of anomaly mediated contributions when $\mu = {\cal O}(m_{3/2})$ and $\tan \beta = {\cal O}(1)$ as expected in the pure gravity mediation model~\cite{PGM} and the minimal split SUSY model~\cite{minimal split}.

In the MSSM, physical masses of the gauginos are obtained by adding the contributions in equations~(\ref{eq: AM}) and (\ref{eq: HT}), and also consider the effect of renormalization group running on the masses down to those scales from $M_{\rm SUSY}$ (the typical mass scale of the scalars). In the left panel of figure~\ref{fig: gaugino masses}, the gaugino masses are shown as a function of $L$ assuming the phase of the Higgsino threshold corrections to be zero (${\rm arg}\,L = 0$), and $M_{\rm SUSY} = m_{3/2} = 100$\,TeV. It can be seen that the coannihilation region of bino and wino is realized when $|L| \sim 3 \, m_{3/2}$.

\subsection{Gaugino masses from a vector-like matter field}
\label{subsec: VM}

When we have a vector-like matter field with its mass of ${\cal O}(m_{3/2})$, there are additional contributions to the gaugino masses. Such a matter field is actually introduced by e.g. models which do not have quantum anomaly in R-symmetry~\cite{Kurosawa:2001iq, Harigaya:2013vja}. The matter field gives the contributions in two ways~\cite{vector matters, Harigaya:2013asa}. First, the field modifies the beta functions, and hence contributes to the gaugino masses via anomaly mediation. Second, as in the case of the Higgsino, threshold corrections from the field exist. Assuming the matter field to be a GUT multiplet to preserve the gauge coupling unification, the sum of these two is parameterized as
\begin{eqnarray}
\Delta M_i^{\rm (VM)} =
\frac{g_i^2}{16\pi^2} \, e^{i\gamma} \, N_{\rm eff} \, m_{3/2},
\label{eq: VM}
\end{eqnarray}
where $\gamma$ is the relative phase between the anomaly mediated contributions and the threshold corrections, which is in fact determined by the phase between the SUSY invariant mass term and the SUSY breaking mass term (non-diagonal one) for the scalar component of the vector-like matter field. The coefficient $N_{\rm eff}$, which can take any real number, depends on the mass terms of the vector-like matter field and is proportional to its Dynkin index~\cite{vector matters, Harigaya:2013asa}.

\begin{figure}[t]
\begin{center}
\includegraphics[width=0.47\linewidth]{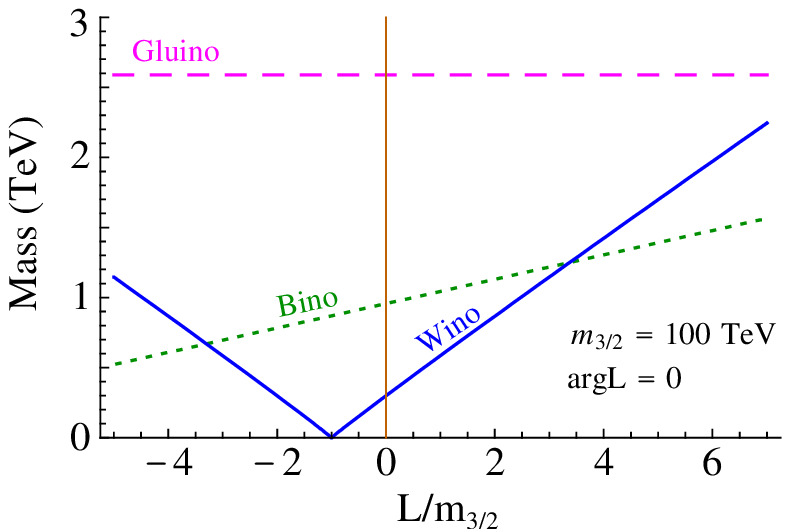}
~~~~
\includegraphics[width=0.47\linewidth]{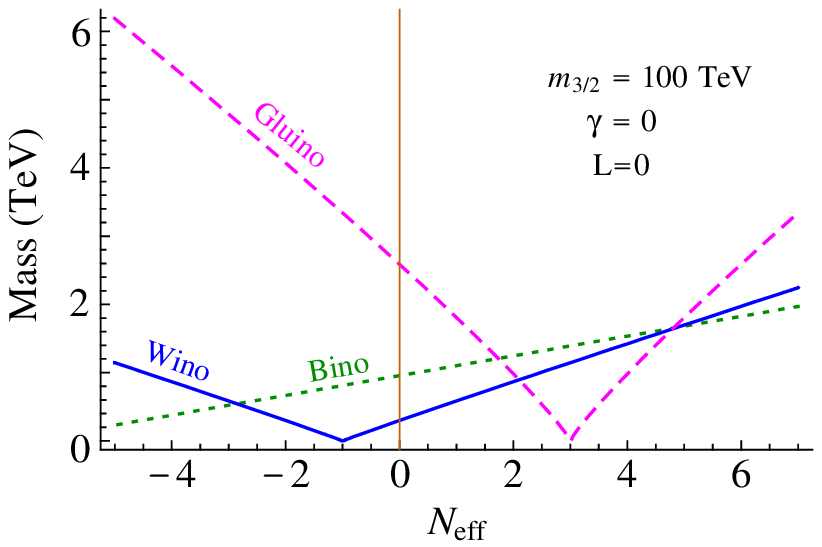}
\caption{\sl \small {\bf Left panel:} Gaugino (bino, wino, and gluino) masses in the high-scale SUSY breaking scenario of the MSSM. {\bf Right panel:} Gaugino masses in the high-scale SUSY breaking scenario of the MSSM plus a vector-like matter field.}
\label{fig: gaugino masses}
\end{center}
\end{figure}

In the right panel of figure~\ref{fig: gaugino masses}, physical masses of the gauginos are shown as a function of $N_{\rm eff}$. The Higgsino threshold corrections and the phase are assumed to vanish ($L = \gamma = 0$) with being $M_{\rm SUSY} = m_{3/2} = 100$\,TeV. It can be seen that the coannihilation region between bino and wino is realized when $N_{\rm eff} \sim -3$, while we have the region between wino and gluino when $N_{\rm eff} \sim 2$. It is even possible to find the region that all gauginos are degenerate when $N_{\rm eff} \sim 5$.

\subsection{Gaugino masses from a Peccei-Quinn sector}
\label{subsec: PQ}

The PQ mechanism~\cite{PQ} has been proposed to solve the strong CP problem~\cite{strong CP}. When a KSVZ-type model~\cite{KSVZ} is adopted, it also gives contributions to the gaugino masses~\cite{Nakayama:2013uta}. Below the PQ breaking scale ($\Lambda_{\rm PQ}$), only the axion multiplet ($a$) is the light degree of freedom in the PQ sector, where the imaginary part of its scalar component is the axion. Because the PQ symmetry is anomalous, the multiplet couples to a gauge kinetic function ($W^\alpha W_\alpha$) as follows:
\begin{eqnarray}
\frac{1}{16} \int {\rm d}^2 \theta \,
(\frac{1}{g^2} + i \frac{\theta_{\rm YM}}{8\pi^2}
+ \frac{N_{\rm DW}}{8\pi^2} \frac{a}{\Lambda_{\rm PQ}}) \,
W^\alpha W_\alpha.
\end{eqnarray}
$\theta_{\rm YM}$ is the theta angle and $N_{\rm DW}$ is the number of domain walls in the PQ sector. The axion multiplet has no supersymmetric potential and hence obtains an F-term of $F_a = {\cal O}(m_{3/2} \Lambda_{\rm PQ})$ due to supergravity effects, which yields\footnote{Corrections by a flat direction coupling to a matter field were first discussed in reference~\cite{Pomarol:1999ie}.}
\begin{eqnarray}
\Delta M_i^{\rm (PQ)} =
\frac{g_i^2}{16\pi^2} N_{\rm DW} \frac{F_a}{\Lambda_{\rm PQ}}.
\end{eqnarray}

In some simple models~\cite{ Nakayama:2013uta, Harigaya:2013asa}, there is no phase degree of freedom in the PQ sector, and hence the contributions $\Delta M_i^{\rm (PQ)}$ are parameterized as those in equation~(\ref{eq: VM}) with $\gamma$ being zero. Physical masses of the gauginos can therefore be read off from the right panel of figure~\ref{fig: gaugino masses}. In models with large $N_{\rm DW}$ such as the one discussed in reference~\cite{Harigaya:2013vja}, the contributions become very significant.

\section{Thermal relic abundance}
\label{sec: abundance}

Before going to discuss the thermal relic abundance of the gaugino dark matter, we write down the low-energy effective lagrangian of the high-scale SUSY breaking scenario at the scale around the gaugino masses. As already mentioned in introduction, the Higgsino is assumed to be much heavier than the gauginos, 
and thus the mixing between bino and wino is approximately given by $m_Z^2/(\mu |\Delta M|) \simeq 10^{-2}(\mu / 100\,{\rm TeV})^{-1}(|\Delta M|/10\,{\rm GeV})^{-1}$ where $\Delta M$ is the mass difference between bino and wino. Even if bino and wino (whose masses are ${\cal O}(10^{2-3})$ GeV) are nearly degenerate, the mixing is less than ${\cal O}(1)\,\%$ in the parameter region of  interest. Therefore, their mass eigenstates are well approximated by their weak eigenstates \footnote{Note that the mixing is significant if Higgsino is light, which is discussed in  reference \cite{Baer:2005zc}. If the sign of $M_1$ and $M_2$ is opposite, bino and wino do not mix each other \cite{Baer:2005jq}}.
The lightest and the second lightest neutralinos are then pure neutral gauginos, while the lightest chargino is the pure charged wino. In following discussion, we denote bino, neutral wino, charged wino, and gluino fields as $\widetilde{B}$, $\widetilde{W}^0$, $\widetilde{W}^-$, and $\widetilde{G}^a$ with $M_1$, $M_2$, $M_2^c$, and $M_3$ being their physical masses, respectively. The mass difference between charged and neutral winos is generated by a quantum correction of the standard model (SM)~\cite{wino mass difference 1}, and has been calculated at two-loop level~\cite{wino mass difference 2}. When the wino mass $|M_2|$ is much larger than the electroweak scale, the difference is about 170\,MeV without depending on $M_2$.

The effective lagrangian involves SM interactions, renormalizable interactions of the gauginos which play important roles to calculate their annihilation cross sections, and higher-dimensional interactions obtained by integrating out heavy fields with masses of ${\cal O}(m_{3/2})$ (sfermions, Higgsino, heavy Higgs bosons):
\begin{eqnarray}
{\cal L}_{\rm eff} &=& {\cal L}_{\rm SM} + {\cal L}_{\rm bino}
+ {\cal L}_{\rm wino} + {\cal L}_{\rm gluino} + {\cal L}_{\rm H.O.},
\label{eq: effective lagrangian} \\
{\cal L}_{\rm bino} &=& (1/2) \,
\overline{\widetilde{B}} (i\slashed{\partial} - M_1) \widetilde{B}, \\
{\cal L}_{\rm wino} &=& (1/2) \,
\overline{\widetilde{W}^0} (i\slashed{\partial} - M_2) \widetilde{W}^0
+ \overline{\widetilde{W}^-} (i\slashed{\partial} - M_2^c) \widetilde{W}^-
\nonumber\\
&& - g \, \overline{\widetilde{W}^-}
\left(s_W \slashed{A} - c_W \slashed{Z} \right)
\widetilde{W}^-
- g \, (\overline{\widetilde{W}^-} \slashed{W}^- \widetilde{W}^0 + h.c. ), \\
{\cal L}_{\rm gluino} &=& (1/2) \,
\overline{\widetilde{G}^a} (i\slashed{\partial} - M_3) \widetilde{G}^a
+ i \, (g_s/2) \, f^{abc}
\overline{\widetilde{G}^a} \slashed{G}^b \widetilde{G}^c.
\end{eqnarray}
$\slashed{A}$, $\slashed{W}^-$, and $\slashed{Z}$ are photon, $W$, and $Z$ boson fields, while $g$, $g_s$, and $s_W = \sin \theta_W$ ($c_W = \cos \theta_W$) are gauge coupling constants of $SU(2)_L$, $SU(3)_c$, and the sine (cosine) of the Weinberg angle, respectively. The SM lagrangian is denoted by ${\cal L}_{\rm SM}$. The last term ${\cal L}_{H.O.}$ involves higher-dimensional interactions: e.g. four Fermi interactions including two gauginos and two SM fermions. The operators play important roles to maintain chemical equilibrium between the lightest and next lightest supersymmetric particles during the coannihilation period via decay, inverse decay, and conversion processes. Since detailed forms of the higher-dimensional interactions are not important for our discussion, we omit to write down those explicitly.

As already mentioned in previous section, we treat gaugino masses ($M_1$, $M_2$, and $M_3$) as free parameters. We thus consider the following four coannihilations below: bino-gluino coannihilation, wino-gluino coannihilation, bino-wino coannihilation, and the coannihilation in which all gauginos participate. The last case is discussed using the model presented in section~\ref{subsec: PQ}, for simplicity.

\subsection{Bino-gluino coannihilation}
\label{subsec: bino-gluino}

It is known that the thermal relic abundance of the dark matter with coannihilation processes is obtained by solving the following Boltzmann equation~\cite{Griest:1990kh}:
\begin{eqnarray}
\frac{dY}{dx} = -\frac{\langle \sigma_{\rm eff}\,v \rangle}{H \, x}
\left( 1 - \frac{x}{3g_{*s}}\frac{dg_{*s}}{dx} \right)
s \, (Y^2-Y^2_{eq}).
\label{eq: Boltzmann}
\end{eqnarray}
$Y$ is the dark matter yield defined by the ratio between the number density of the dark matter particle and the entropy density of the universe, $s = g_{*s}\,(2\pi^2/45)(m^3/x^3)$, with $x$ being the inverse temperature of the universe in unit of the dark matter mass, $x=m/T$. The Hubble parameter $H$ and the equilibrium yield $Y_{eq}$ are given by $H = (g_*/90)^{1/2}(\pi/m_{\rm pl})(m^2/x^2)$ and $Y_{eq} = (g_{\rm eff}\,m^3/s)\,x^{-3/2}\,e^{-x}/(2\pi)^{3/2}$, respectively, where $m_{\rm pl}$ is the reduced Planck mass. The massless degrees of freedom for energy and entropy are denoted by $g_*$ and $g_{*s}$, respectively, and those are evaluated according to reference~\cite{massless degrees} using lattice data of the QCD phase transition~\cite{Karsch:2000ps}. The effective annihilation cross section $\sigma_{\rm eff}$ is then given by
\begin{eqnarray}
\sigma_{\rm eff}\,v = \sum_{i,\,j} (\sigma_{ij}\,v) \,
\frac{g_i\,g_j}{g^2_{\rm eff}}
(1 + \Delta_i)^{3/2}(1 + \Delta_j)^{3/2}
\exp[-x\,(\Delta_i + \Delta_j)],
\label{eq: effective annihilation}
\end{eqnarray}
where $\sigma_{ij}$ is the annihilation cross section between particles `$i$' and `$j$' with $g_i$ and $g_j$ being their spin (color) degrees of freedom, $v$ is the relative velocity between the particles, and $g_{\rm eff}$ is the effective degree of freedom for `dark matter particles', $g_{\rm eff} = \sum_{i} g_i\,(1+\Delta_i)^{3/2}\exp[-x\Delta_i]$ with $\Delta_i = (m_i - m)/m$. The mass of the particle `$i$' is denoted by $m_i$, and $m_1 = m$ corresponds to the dark matter mass. The cross section with the bracket, $\langle \sigma_{\rm eff}\,v \rangle$, in equation (\ref{eq: Boltzmann}) represents the one which is averaged by the dark matter velocity distribution at the temperature $T$.

For the case of bino-gluino coannihilation in the high-scale SUSY breaking scenario, the annihilations of $\widetilde{B} \, \widetilde {B} \to$ SMs and $\widetilde{B} \, \widetilde{G} \to$ SMs are suppressed due to heavy sfermions and Higgsinos. Only the annihilation $\widetilde{G} \, \widetilde{G} \to$ SMs thus contributes to the effective annihilation cross section. It is worth noting here that the chemical equilibrium between coannihilating particles during the freeze-out epoch is maintained thanks to higher-dimensional operators in the lagrangian (\ref{eq: effective lagrangian}): the (inverse) decay rate of the gluino and the conversion rate between bino and gluino are enough larger than the expansion rate of the universe $H$, so that the ratio of number densities between the coannihilating particles is determined only by the temperature $T$. In the gluino annihilation, the Sommerfeld effect may enhance or suppress its cross section~\cite{sommerfeld effect}. The effect can be interpreted as the one distorting wave-functions of incident particles due to long-range force acting between them, and it is incorporated through the following formula at leading order
\footnote{Coannihilation between gluino and neutralino (corresponding to bino and wino in our case) has  been already considered in references~\cite{Profumo:2004wk,Feldman:2009zc} without including the Sommerfeld effect,
while the case of gluino being LSP is studied with including the Sommerfeld effect \cite{Baer:1998pg}.
The effect has very recently involved in the coannihilation between bino ($SU(2)_L$-singlet) and gluino ($SU(3)_c$-octet) in reference~\cite{deSimone:2014pda}.}:
\begin{eqnarray}
\sigma v = (\sigma_0 v) \times \lim_{r \to \infty}|\psi(r)|^2,
\label{eq: gluino annihilation}
\end{eqnarray}
where $\sigma_0$ is the self-annihilation cross section of the gluino calculated in a usual perturbative way, while $|\psi(r)|^2$ is so-called the Sommerfeld factor. The factor is calculated by solving the following Shr${\rm \ddot{o}}$dinger equation,
\begin{eqnarray}
\left[-\frac{1}{M_3} \frac{d^2}{dr^2} + V(r) \right] \psi(r) = E \, \psi(r),
\end{eqnarray}
with the boundary condition: the wave-function $\psi(r)$ has only an out-going wave at $r \to \infty$ with its normalization fixed to be $\psi(0) = 1$.

Potential $V(r)$ in the above Shr${\rm \ddot{o}}$dinger equation depends on which color representation the incident gluino pair has. The product of two color adjoint representations is decomposed into $1 \oplus 8_A \oplus 8_S \oplus 10 \oplus \overline{10} \oplus 27$. With the fact that the s-wave process dominates the annihilation and the gluino is a Majorana fermion, the representations $1$, $8_S$, and $27$ must form spin-0 states, while other representation $8_A$, $10$, and $\overline{10}$ must form spin-1 states. The potential is then given by
\begin{eqnarray}
V_R(r) \simeq c_R \, \alpha_s/r,
\end{eqnarray}
with the coefficient $c_R = -3$, $-3/2$, $-3/2$, $0$, $0$, and $1$ for representations $1$, $8_S$, $8_A$, $10$, $\overline{10}$, and $27$, respectively. It then turns out that the potential gives repulsive force for the representation $27$, and its annihilation cross section is highly suppressed. For the representations $10$ and $\overline{10}$, the potential vanishes, and their initial wave-functions are not distorted. For the representations $1$, $8_S$, and $8_A$, the potential gives attractive force, and their annihilation cross sections are expected to be enhanced. In fact, the Shr${\rm \ddot{o}}$dinger equation can be solved analytically when $V(r)$ is approximated by the Coulomb potential, and the Sommerfeld factor becomes
\begin{eqnarray}
|\psi(r)|^2 = 
\frac{2\pi c_R \alpha_s/v}{\exp [2\pi c_R \alpha_s/v] - 1},
\end{eqnarray} 
with $v$ being the relative velocity between the incident gluino pair. The factor is actually enhanced by $1/v$ for a negative $c_R$, while it is suppressed for a positive $c_R$. Here, we should mention about which energy scale we should use to evaluate $\alpha_s$ in the factor, because higher order QCD corrections to $V(r)$ depends significantly on the scale. According to the prescription in reference~\cite{Nagano:1999nw}, we take the scale $\mu$ obtained by solving the flowing self-consistency equation:
\begin{eqnarray}
\mu &=& (M_3/2) \, |c_R| \, \alpha_s(\mu).
\end{eqnarray} 
In order to evaluate the factor more accurately, we should calculate the potential including higher order QCD corrections as well as finite temperature corrections, because the freeze-out phenomena occurs before QCD phase transition (say, in the symmetric phase), which is postponed to future work.

Because the Sommerfeld effect depends on the representation of the incident gluino pair, the annihilation cross section, $\sigma_0\,v$, in equation (\ref{eq: gluino annihilation}) must be calculated in each representation. The cross section is given by
\begin{eqnarray}
\sigma_0\,v|_{R = 1~\,} &=& 4 \pi \alpha_s^2 c_R^2/M_3^2, \\
\sigma_0\,v|_{R = 8_S} &=& 4 \pi \alpha_s^2 c_R^2/M_3^2, \\
\sigma_0\,v|_{R = 8_A} &=&
 (\pi \alpha_s^2/M_3^2) {\textstyle \sum_f}
(2 + m_f^2/M_3^2) (1 - m_f^2/M_3^2)^{1/2}, \\
\sigma_0\,v|_{R = 27 \,} &=& 4 \pi \alpha_s^2 c_R^2/M_3^2,
\end{eqnarray}
while the cross sections for the representations $10$ and $\overline{10}$ vanish. The cross section for $R = 8_A$ comes from annihilations to various quark pairs, while those for other representations ($R = 1$, $R = 8_S$, and $R = 27$) are from annihilation to gluon pair. As a result, the contribution to the effective annihilation cross section in equation (\ref{eq: effective annihilation}) from the gluino self-annihilation is given by
\begin{eqnarray}
\sigma_{\tilde{G}\tilde{G}} \, v = (1/256)
(\sigma\,v|_{R = 1} + 8\,\sigma\,v|_{R = 8_S}
+ 3 \times 8\,\sigma\,v|_{R = 8_A} + 27\,\sigma\,v|_{R=27}),
\end{eqnarray}
which is consistent with reference \cite{Baer:1998pg}. 

\begin{figure}[t]
\begin{center}
\includegraphics[scale=.77]{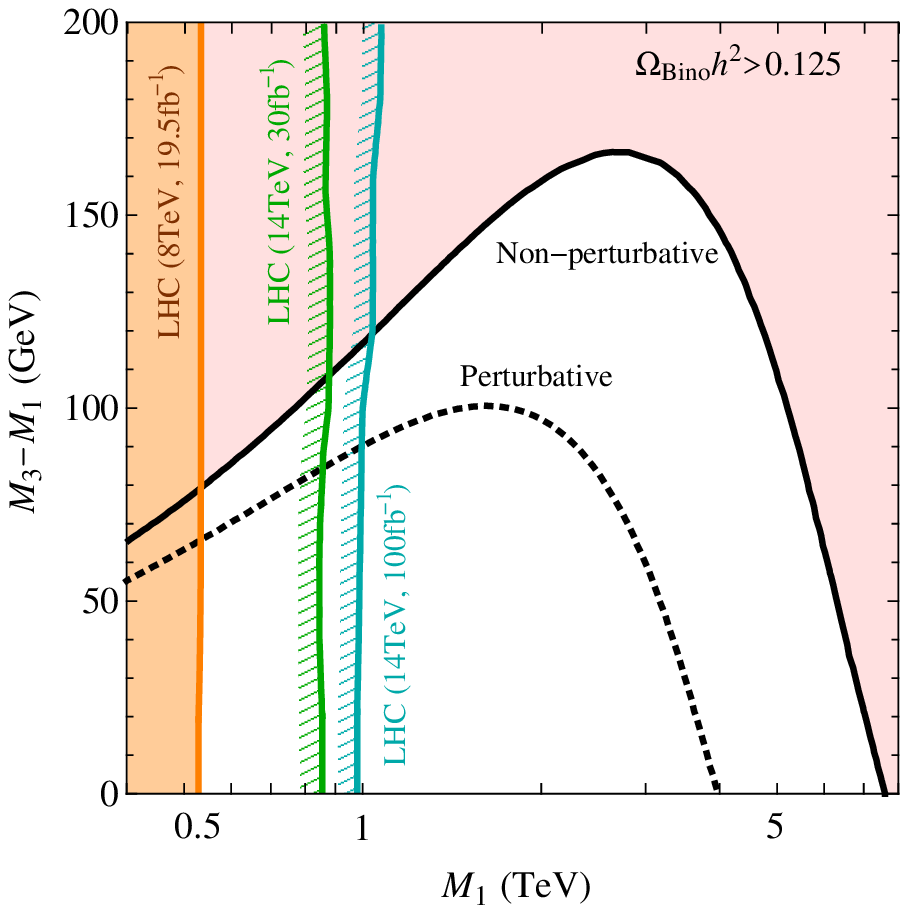}
~~
\includegraphics[scale=.55]{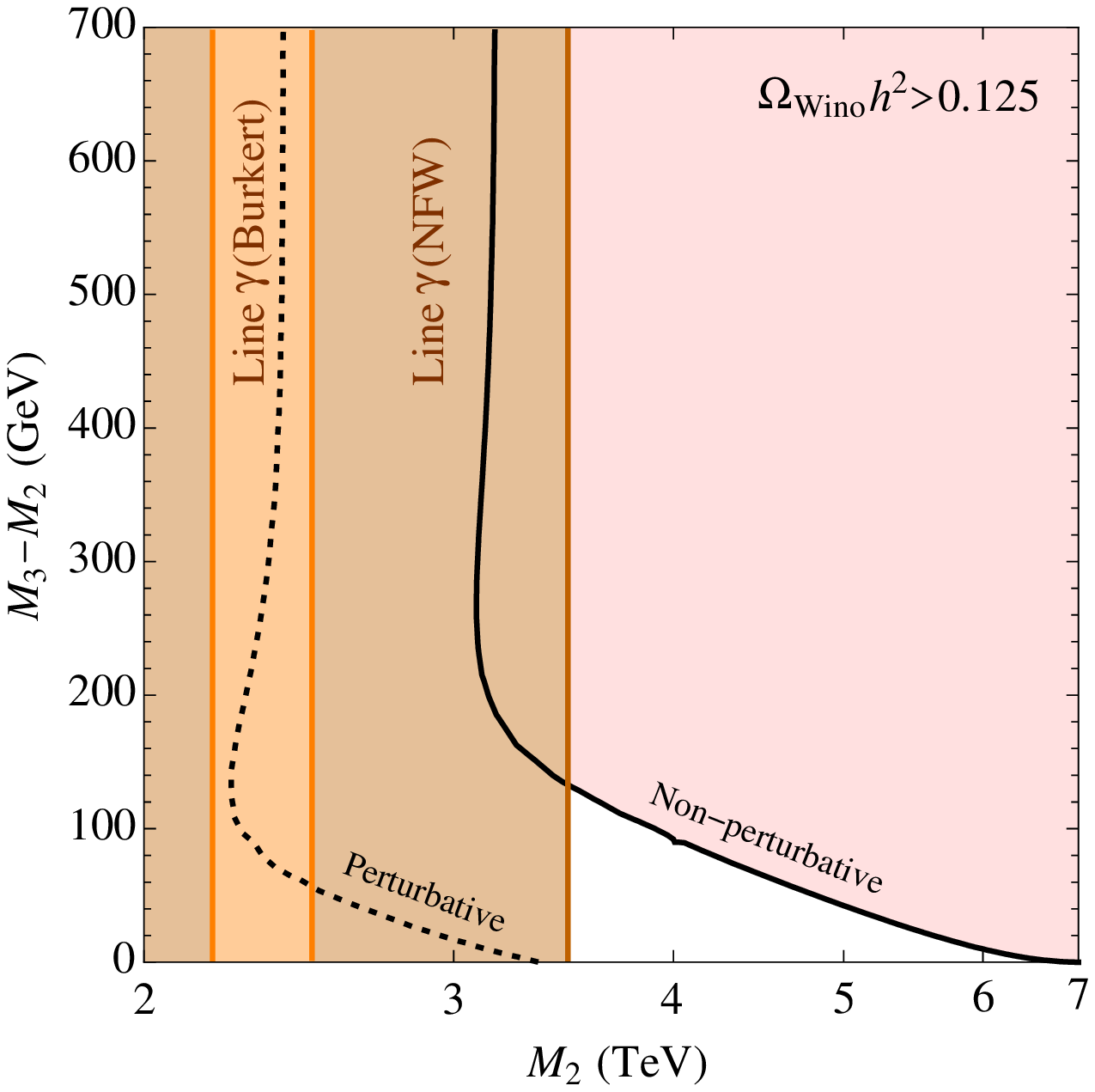}
\caption{\sl \small {\bf Left panel:} Coannihilation region between bino and gluino. The bino dark matter is over-produced in the region above the black line. For comparison, the result without the Sommerfeld effect is shown as the black dotted line. Current and future-expected limits on the region from the LHC experiment are also shown. {\bf Right panel:} Coannihilation region between wino and gluino. The black solid and dotted lines have the same meanings as those of the left panel. A limit on the wino dark matter obtained from the monochromatic line-gamma ray search (by observing the galactic center) at the H.E.S.S. experiment is also shown. See text for more details.}
\label{fig: gluino coannihilations}
\end{center}
\end{figure}

With the annihilation cross section discussed above and solving the Boltzmann equation (\ref{eq: Boltzmann}), we obtain the final yield of the dark matter particle, $Y(\infty)$. The thermal relic abundance of the dark matter is then given by $\Omega h^2 = m\,s_0\,Y(\infty)/(\rho_c\,h^{-2})$ with $s_0=2889$\,cm$^{-3}$ and $\rho_c\,h^{-2} = 1.054 \times 10^{-5}$\,GeV\,cm$^{-3}$. In the left panel of figure~\ref{fig: gluino coannihilations}, the coannihilation region of bino and gluino is shown. Along the black solid line, the resultant abundance coincides with the observed upper limit, $\Omega^{\rm (obs.)} h^2 = 0.125$. The region below (above) the line, the abundance is smaller (larger) than the value. As a reference, we have shown the region obtained neglecting the Sommerfeld effect~\cite{Profumo:2004wk,Feldman:2009zc}, which is denoted by the black dotted line. It can be seen that the bino dark matter can be as heavy as 7--8\,TeV due to the coannihilation.
In the plot, current~\cite{Chatrchyan:2014lfa} and future-expected~\cite{Bhattacherjee:2013wna} limits on the region obtained from the LHC experiment are also shown as orange and green/blue solid lines, respectively. The current limit is obtained by 19.5\,fb$^{-1}$ data at 8\,TeV running, while future-expected limits are assuming 30 and 100\,fb$^{-1}$ data at 14\,TeV running. Search for the gluino, which is degenerated with a neutralino (bino) with the mass difference of ${\cal O}(100)$\,GeV, is therefore mandatory to explore the gluino-bino coannihilation region of the high-scale SUSY breaking scenario. In this search, the gluino pair production associated with the initial state radiation (ISR) gluon(s) will play an important role.

\subsection{Wino-gluino coannihilation}
\label{subsec: wino-gluino}

Calculation of the dark matter abundance in wino-gluino coannihilation region is essentially the same as that in previous subsection. Only the difference is that annihilations of wino dark matter and its $SU(2)_L$ partners also contribute to the effective annihilation cross section (\ref{eq: effective annihilation}). The coannihilation between wino and gluino is again suppressed because of heavy sfermions and Higgsinos. In the wino annihilations, there are six annihilation modes: $\widetilde{W}^0 \widetilde{W}^0$, $\widetilde{W}^+ \widetilde{W}^-$, $\widetilde{W}^0 \widetilde{W}^\pm$, and $\widetilde{W}^\pm \widetilde{W}^\pm$. Remembering the fact that the neutral wino is a Majorana fermion, initial states of $\widetilde{W}^0 \widetilde{W}^0$ and also $\widetilde{W}^\pm \widetilde{W}^\pm$ form only spin-0 states. Initial states of other modes, on the other hand, form both spin-0 and spin-1 states. See appendix~\ref{app: wino annihilations} for concrete expressions of their annihilation cross sections. As in the gluino annihilation, the wino annihilations also receive the Sommerfeld effect. In the annihilations, the potentials $V(r)$ in their Schr$\ddot{\rm o}$dinger equations are generated by exchanging photons (Coulomb potential) and $W/Z$ bosons (Yukawa potential) between the incident particles. Since the Sommerfeld effect on the annihilations have already been discussed in the literature~\cite{sommerfeld effect}, we omit to write down those explicitly.

The coannihilation region between wino and gluino is shown in the right panel of figure~\ref{fig: gluino coannihilations}. 
The relic abundance of neutral wino is below the observed upper limit on the left side of the black solid line, when the Sommerfeld effect is included. 
For comparison, the result without the Sommerfeld effect is shown by the black dotted line.
At the right ends of the lines, gluino and wino are almost degenerated with each others.
In this case, due to the large annihilation cross section of gluino in comparison with that of wino, the dark matter abundance is essentially determined by the annihilation cross section of gluino (so-called Profumo-Yaguna formula~\cite{Profumo:2004wk}).
It can be seen that the wino can be as heavy as 7\,TeV because of the coannihilation. When the mass difference between wino and gluino is large enough, the solid line asymptotically approaches $M_2 \simeq 3.1$\,TeV, which is the mass predicted by the usual wino dark matter. A bumpy structure can be seen on the black solid (dotted) line at $M_3 - M_2 \sim 200$\,GeV ($M_3 - M_2 \sim 100$\,GeV), which originates in the gluino contribution; it is somewhat suppressed by the Boltzmann factor in this region and its annihilation cross section becomes comparable to the wino's, leading to the suppression of the effective annihilation cross section due to the increase of $g_{\rm eff}$. Another limit on the wino dark matter is also shown in the plot, which is obtained from the monochromatic gamma-ray search (by observing the galactic center) at the H.E.S.S. experiment~\cite{Abramowski:2013ax}. The limit depends strongly on the dark matter profile at the center. The orange region is the limit adopting the NFW (cuspy) profile~\cite{Navarro:1995iw}, while the brown region is the one adopting the Burkert (cored) profile~\cite{Burkert:1995yz}. The limits are estimated with allowing 2$\sigma$-deviation from circular velocity data of our galaxy~\cite{Nesti:2013uwa}. It is interesting to see that, even if we take the limit adopting the NFW (cuspy) profile, we can find the parameter region consistent with cosmology (the thermal relic abundance of the dark matter).

\subsection{Bino-wino coannihilation}
\label{subsec: bino-wino}

Calculation of the dark matter abundance in bino-wino coannihilation region is also the same as those in previous subsections. In this region, only the wino annihilations contribute to the effective annihilation cross section (\ref{eq: effective annihilation}). Other annihilation processes between binos and between bino and wino are suppressed again because of heavy sfermions and Higgsinos. Since both bino and wino can be a dark matter in this coannihilation, we discuss the two cases separately.

\begin{figure}[t]
\begin{center}
\includegraphics[scale=0.76]{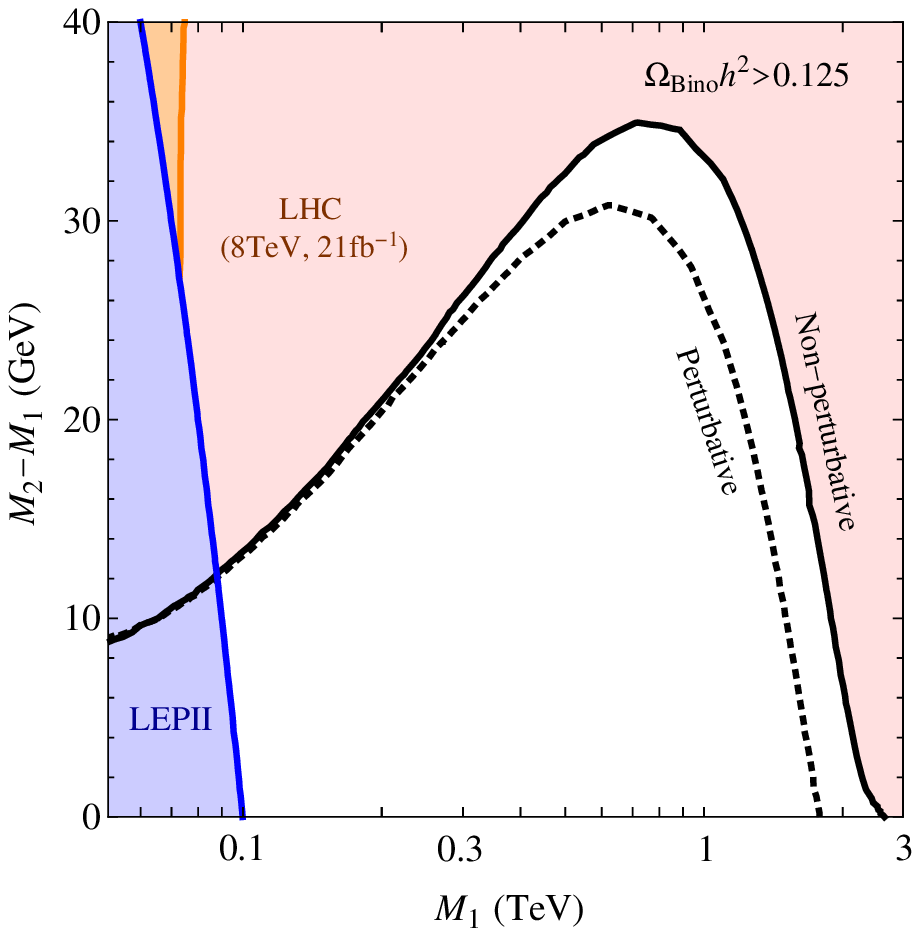}
~~
\includegraphics[scale=0.785]{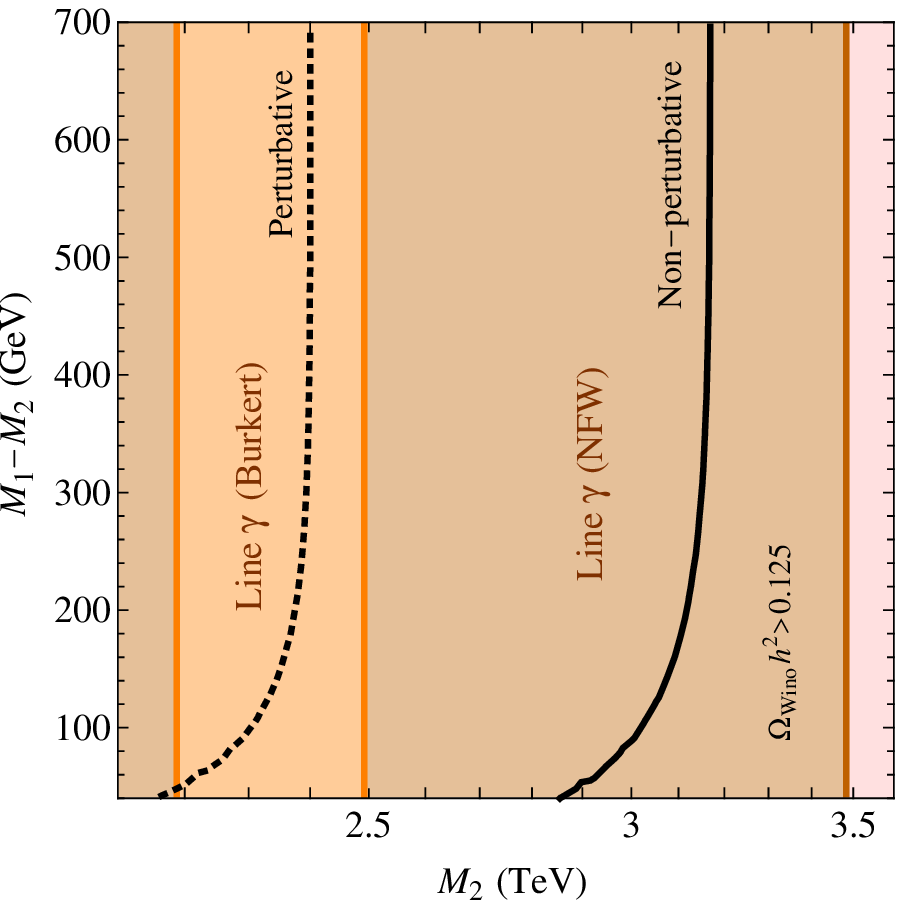}
\caption{\sl \small {\bf Left panel:} Coannihilation region between bino and wino with the bino being dark matter. Black solid and dotted lines have the same meanings as those of previous figures. Limits from the LEP\,II and LHC experiments are also shown as blue and orange lines. {\bf Right panel:} Coannihilation region between bino and wino with the wino being dark matter. The black solid and dotted lines have the same meaning as those of the left panel. A limit on the wino dark matter obtained by the monochromatic line-gamma ray search (observing the galactic center) at the H.E.S.S. experiment is also shown.}
\label{fig: bino-wino coannihilations}
\end{center}
\end{figure}

Bino-wino coannihilation with the bino being dark matter is similar to bino-gluino coannihilation, as seen in the left panel of figure~\ref{fig: bino-wino coannihilations}. Black solid and dotted lines have the same meanings as those of previous figures. The bino dark matter can be as heavy as 3\,TeV due to the coannihilation. We have also shown other limits obtained by collider physics. The blue region has been excluded by the LEP\,II experiment, in which the wino pair production was searched for via a radiative return process~\cite{Heister:2002mn}. The orange region has been ruled out by the LHC (ATLAS) experiment, in which the $\widetilde{W}^\pm \widetilde{W}^0$ production was searched for via its decay into three leptons, $\widetilde{W}^\pm \widetilde{W}^0 \to (W^\pm \widetilde{B}) \, (Z \widetilde{B}) \to \nu \, \ell^\pm \, \ell^+ \, \ell^- \, \widetilde{B} \, \widetilde{B}$~\cite{Aad:2014nua}. In the analysis, charged and neutral winos are assumed to decay into off-shell $W$ and $Z$ bosons with 100\% ratio. In the high-scale SUSY breaking scenario, almost all charged winos actually decay into $W^*$, for it is governed by a dimension-five operator involving neutral wino, bino, and two Higgs doublets, which causes the transition from neutral wino into bino after the decay $\widetilde{W}^+ \to \widetilde{W}^0 W^*$. Other decay channels of the charged wino are from dimension-six operators. On the other hand, this dimension-five operator also induces the decay of the neutral wino into a off-shell Higgs boson, though its fraction is suppressed by the bottom Yukawa coupling. Decay channels of the neutral wino into two leptons are from dimension-six operators. The neutral wino thus decays mainly into two leptons when Higgsinos or sleptons are somewhat lighter than other heavy SUSY particles. It turns from the figure that detecting soft leptons of ${\cal O}(10)$\,GeV at 14\,TeV running will play a crucial role to explore this coannihilation region.

As already mentioned, the mixing between bino and wino is negligibly small in the parameter region of interest since we assume that Higgsino is much heavier than gauginos.
For instance, the mixing is ${\cal O}(1) \,\%$ when $M_1\simeq 100$\,GeV and $M_2-M_1\simeq 10$\,GeV, which is too small and  irrelevant in our result.
For the heavy bino case, the mixing becomes around $10 \,\%$ when $M_1\simeq 3$\,TeV and $M_2-M_1\simeq {\cal O}(1)$\,GeV, which we have neglected in the analysis.

The coannihilation between bino and wino with the wino being dark matter is, on the other hand, similar to wino-gluino coannihilation, which is shown in the right panel of figure~\ref{fig: bino-wino coannihilations}. Black solid and dotted lines have the same meanings as before. The difference between the coannihilations can be seen at the region that coannihilating particles are highly degenerated in mass. In wino-gluino coannihilation, the effective annihilation cross section is enhanced by the gluino annihilation at this region, while it is suppressed by very small (almost zero) annihilation of bino in bino-wino coannihilation. As a result, the wino mass coinciding with the observed upper limit is decreased to 2.8\,TeV, which is smaller than the mass predicted by the usual wino dark matter, $M_2 \simeq 3.1$\,TeV. When the mass difference between bino and wino is large enough, the black solid line approaches this value.
In the plot, a limit from the H.E.S.S. experiment is also shown as in the case of wino-gluino coannihilation. It then turns out that, if we take the limit adopting the NFW profile, all region is excluded, though the use of the NFW profile seems too aggressive to conclude that the coannihilation region has completely been ruled out.

\subsection{Coannihilation in which all gauginos participate}
\label{subsec: bino-wino-gluino}

\begin{figure}[t]
\begin{center}
\includegraphics[scale=0.80]{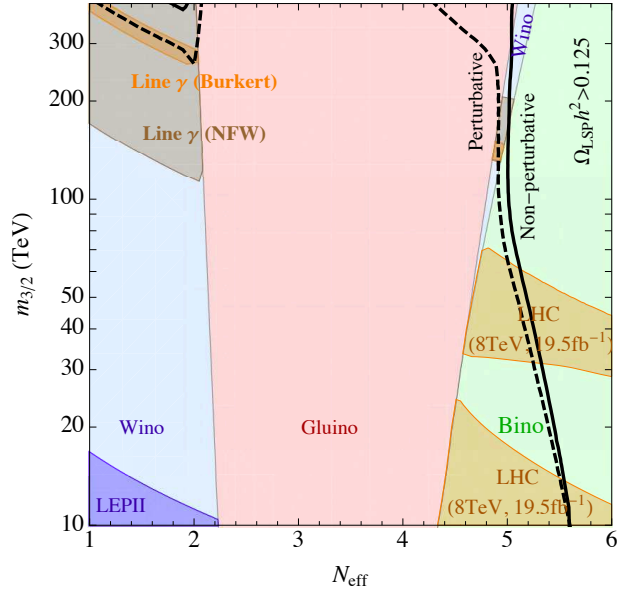}
\caption{\sl \small Coannihilation region in which all gauginos participate in the framework of the Peccei-Quinn extension of the MSSM. The Higgsino threshold corrections are assumed to vanish ($L = 0$). Black solid and dotted lines have the same meaning as before. Wino, gluino, and bino are the lightest supersymmetric particle in regions painted by light blue, light red, and light green, respectively. Several experimental constraints are also shown by the same colors as those of previous figures.}
\label{fig: all gauginos}
\end{center} 
\end{figure}

Here, we consider the thermal relic abundance of dark matter in the framework of the Peccei-Quinn extension of the MSSM, as an example with coannihilation in which all gauginos participate. We have scanned the following parameter region: 10\,TeV $< m_{3/2} <$ 400\,TeV and 1 $< N_{\rm DW} <$ 6 with the Higgsino threshold corrections $L$ being neglected. See also figure~\ref{fig: gaugino masses} for your reference, where $N_{\rm DW}$ can be regarded as $N_{\rm eff}$ with $\gamma$ being zero. The result is shown in figure~\ref{fig: all gauginos} as a function of $m_{3/2}$ and $N_{\rm DW}$. Black solid and dotted lines are the same as those in previous figures. Roughly speaking, the result is divided into three regions: the wino dark matter region (painted by blue) when $N_{\rm DW}$ is smaller than about two, the bino dark matter region (painted by green) when $N_{\rm DW}$ is larger than 4--5, and the gluino dark matter region (painted by pink) in between. The gluino dark matter region is, of course, strongly disfavored by various experiments and observations.

It can be seen from the figure that the coannihilation region in which all gauginos participate is realized at $N_{\rm DW} \sim 5$ and $m_{3/2} > 50$\,TeV, where the region appears as a narrow blue band sandwiched by pink and green regions. It is also worth pointing out that the black solid line is across the region. This fact means that there is a region providing the correct relic abundance due to coannihilation of all gauginos: the region is found to be $N_{\rm eff} \simeq 5$ and $m_{3/2} \simeq 250$\,TeV, corresponding to $(M_1, M_2, M_3) \simeq$ (4250\,GeV ,4210\,GeV, 4220\,GeV). In the plot, we have also shown several constraints from LEP\,II, LHC, and H.E.S.S. experiments as in cases of coannihilation regions previously discussed. It can be seen that the coannihilation region mentioned above, namely $N_{\rm eff} \simeq 5$ and $m_{3/2} \simeq 250$\,TeV, evades all of the constraints.

\section{Summary}
\label{sec: summary}

In this article, we have calculated the thermal relic abundance of the gaugino (bino or wino) dark matter in the framework of the high-scale SUSY breaking scenario, with especially focusing on various coannihilation regions between gauginos. Sommerfeld effects on wino and gluino annihilations have been involved in our calculations, which are known to give significant contributions to their annihilation cross sections. Based on obtained results, we have also discussed some implications to gaugino searches at collider and indirect detection experiments of dark matter.

For bino-gluino coannihilation, the bino dark matter can be as heavy as 7--8\,TeV due to the coannihilation. The mass difference between gluino and bino is required to be ${\cal O}(100)$\,GeV. In order to explore the coannihilation region of the high-scale SUSY breaking scenario, the search for the gluino degenerated with the lightest neutralino at the future LHC experiment is mandatory. The gluino pair production associated with ISR gluon(s) will play an important role. Notice that the gluino lighter than 500\,GeV has already been excluded by current LHC data.

The wino-gluino coannihilation region has rich phenomenology. The wino can be as heavy as 6--7\,TeV, which is much heavier than 3.1\,TeV for which the wino dark matter abundance coincides with the observed value without the coannihilation. Search for monochromatic gamma-rays whose energy is larger than 3.1\,TeV is of particular importance, because the existence of such a line signal indicates the participation of the coannihilation, leading to a crucial implication to collider experiments. If the dark matter profile of our galaxy is cuspy (e.g. NFW profile) at the galactic center, the wino mass of 3--3.5\,TeV is excluded.

In bino-wino coannihilation region, both bino and wino can be dark matter. When the bino is dark matter, the typical mass difference between wino and bino is ${\cal O}(10)$\,GeV, and the dark matter mass is at most less than 3\,TeV. In order to explore the region at 14\,TeV running of the LHC experiment, the search for winos decaying into the bino dark matter by emitting soft leptons of ${\cal O}(10)$\,GeV will be important. The bino mass less than about 100\,GeV has already been excluded by the LEP\,II experiment. When the wino is dark matter, its mass is predicted to be within the range 2.9--3.1\,TeV. The search for the monochromatic gamma-ray is again important to prove the region, as is the case of wino-gluino coannihilation. If the dark matter profile is cuspy (i.e. NFW profile), the wino dark matter in this coannihilation region is excluded by current H.E.S.S. data.

We have also calculated the thermal relic abundance of dark matter in the Peccei- Quinn extension of the MSSM.
We have found that in some portions of the parameter space, coannihilations between all gauginos are important to obtain the observed dark matter abundance.
In this case, gaugino masses are typically degenerated with each others around 4 TeV with the wino being dark matter.
Such high energy region cannot be explored by the 14TeV running of the LHC, but a future 100 TeV collider potentially discovers 4 TeV gluino~\cite{Low:2014cba}.
Search for monochromatic gamma-rays whose energy is larger than 3.1\,TeV is also of particular importance as is the case with the wino-gluino coannihilation.

The degeneration of gluino and bino or wino in their masses cannot be achieved
in the high-scale SUSY breaking scenario involving only the MSSM matter contents~\cite{PGM},
since the gaugino masses are determined solely by the anomaly mediation and the higgsino threshold correction, and hence are strongly restricted.
Once the degenerated spectrum is found, we need an extension of the MSSM such as the Peccei-Quinn mechanism or vector-like extra chiral multiplets.
The discovery of degenerated gauginos would shed light on the extension of the MSSM whose energy scale is much higher than $M_{\rm SUSY}$.

\section*{Acknowledgments}

This work is supported by Grant-in-Aid for Scientific Research from the Ministry of Education, Science, Sports and Culture (MEXT), Japan, Nos. 23740169 \& 22244021 (S.M.) and also by World Premier International Research Center Initiative (WPI Initiative), MEXT, Japan. K.H. and K.K. acknowledge the support by Japan Society of the Promotion of Science (JSPS) Research Fellowship for Young Scientists.

\appendix

\section{Wino annihilations}
\label{app: wino annihilations}

Here, we summarize cross sections, which are used to calculate the contribution from wino annihilations. As already mentioned in main text, there are six annihilation modes: $\widetilde{W}^0 \widetilde{W}^0$, $\widetilde{W}^+ \widetilde{W}^-$, $\widetilde{W}^0 \widetilde{W}^\pm$, and $\widetilde{W}^\pm \widetilde{W}^\pm$. Initial states of $\widetilde{W}^0 \widetilde{W}^0$ and $\widetilde{W}^\pm \widetilde{W}^\pm$ form only spin-0 states, while those of other modes form both spin-0 and spin-1 states. Below, we carefully present the cross sections in each mode .

\subsection{$\widetilde{W}^- \widetilde{W}^-$ annihilation}

Since this is the annihilation between identical particles, its initial sate forms only a spin-0 state, and it annihilates into $W^- W^-$ pair. The cross section shown here can also be applied to its conjugate case, namely $\widetilde{W}^+ \widetilde{W}^+ \to W^+ W^+$.
{\small
\begin{eqnarray}
\sigma_0 v|_{WW} =
\frac{4\pi \alpha_2^2}{M_2^2}
\left[ 1 - \frac{m_W^2}{M_2^2} \right]^{3/2}
\left[ 1 - \frac{m_W^2}{2M_2^2} \right]^{-2}.
\end{eqnarray}
}

\subsection{$\widetilde{W}^0 \widetilde{W}^-$ annihilations}

Cross sections presented here can also be applied their conjugate cases. When the initial state forms a spin-0 state, it annihilates into $W^- Z$ and $W^- \gamma$.
{\small
\begin{eqnarray}
\sigma_0 v|_{WZ} &=&
\frac{2\pi \alpha_2^2 c_W^2}{M_2^2}
\left[ 1 - \frac{m_W^2+m_Z^2}{2M_2^2}
+ \frac{\left( m_W^2 - m_Z^2 \right)^2}{16M_2^4} \right]^{3/2}
\left[ 1 - \frac{m_W^2+m_Z^2}{4M_2^2} \right]^{-2},
\\
\sigma_0 v|_{W\gamma} &=&
\frac{2\pi \alpha_2^2 s_W^2}{M_2^2}
\left[ 1 - \frac{m_W^2}{2M_2^2} + \frac{m_W^4}{16M_2^4} \right]^{3/2}
\left[ 1 - \frac{m_W^2}{4M_2^2} \right]^{-2}.
\end{eqnarray}
}

On the other hand, when the initial state forms a spin-1 state, it annihilates into $f \bar{f}^\prime$, $W^- h$, and $W^- Z$. Those cross sections are given as follows. Below, we only show the cross section of $\widetilde{W}^0 \widetilde{W}^- \to e^- \bar{\nu}$ as a representative of $\widetilde{W}^0 \widetilde{W}^- \to f \bar{f}^\prime$.
{\small
\begin{eqnarray}
\sigma_0 v|_{e\bar{\nu}} &=&
\frac{\pi \alpha_2^2}{3M_2^2}
\left[ 1 - \frac{m_W^2}{4M_2^2} \right]^{-2}
\left[ 1 - \frac{m_e^2}{4M_2^2} \right]^2
\left[ 1 + \frac{m_e^2}{8M_2^2} \right],
\\
\sigma_0 v|_{Wh} &=&
\frac{\pi \alpha_2^2}{12M_2^2}
\left[ 1 - \frac{m_W^2}{4M_2^2} \right]^{-2}
\left[ \left( 1 + \frac{m_W^2-m_h^2}{4M_2^2} \right)^2
+ 2 \frac{m_W^2}{M_2^2} \right]
\nonumber \\ && \times
\left[ 1 - \frac{m_W^2 + m_h^2}{2M_2^2}
+ \frac{ \left( m_W^2 - m_h^2 \right)^2}{16M_2^4} \right]^{1/2},
\\
\sigma_0 v|_{WZ} &=&
\frac{\pi \alpha_2^2}{12M_2^2}
\left[ 1 - \frac{m_W^2+m_Z^2}{2M_2^2}
+ \frac{ \left( m_W^2 - m_Z^2 \right)^2}{16M_2^4} \right]^{3/2}
\left[ 1 - \frac{m_W^2+m_Z^2}{4M_2^2} \right]^{-2}
\nonumber \\ && \times
\left[ 1 - \frac{m_W^2}{4M_2^2} \right]^{-2}
\left[ 1 + \frac{5}{2}\frac{m_W^2+m_Z^2}{M_2^2}
+ \frac{m_W^4 + m_Z^4 + 10 m_W^2 m_Z^2}{16M_2^4} \right].
\end{eqnarray}
}

\subsection{$\widetilde{W}^0 \widetilde{W}^0$ annihilation}

Since the neutral wino is a Majorana particle, its initial state forms only a spin-0 state. A pair of the neutral wino annihilates only into $W^+W^-$.
{\small
\begin{eqnarray}
\sigma_0 v|_{WW} =
\frac{8\pi \alpha_2^2}{M_2^2}
\left[ 1 - \frac{m_W^2}{M_2^2} \right]^{3/2}
\left[ 1 - \frac{m_W^2}{2M_2^2} \right]^{-2}.
\end{eqnarray}
}

\subsection{$\widetilde{W}^+ \widetilde{W}^-$ annihilations}

When the initial state forms a spin-0 state, it annihilates into $\gamma \gamma$, $W^+ W^-$, $Z Z$, and $Z \gamma$. Corresponding cross sections of these annihilation channels are as follows:
{\small
\begin{eqnarray}
\sigma_0 v|_{\gamma \gamma} &=&
\frac{4\pi \alpha_2^2 s_W^4}{M_2^2},
\\
\sigma_0 v|_{WW} &=&
\frac{2\pi \alpha_2^2}{M_2^2}
\left[ 1 - \frac{m_W^2}{M_2^2} \right]^{3/2}
\left[ 1 - \frac{m_W^2}{2M_2^2} \right]^{-2},
\\
\sigma_0 v|_{ZZ} &=&
\frac{4\pi \alpha_2^2 c_W^4}{M^2}
\left[ 1 - \frac{m_Z^2}{M_2^2} \right]^{3/2}
\left[ 1 - \frac{m_Z^2}{2M_2^2} \right]^{-2},
\\
\sigma_0 v|_{Z \gamma} &=&
\frac{8\pi \alpha_2^2 c_W^2 s_W^2}{M^2}
\left[ 1 - \frac{m_Z^2}{2M_2^2} + \frac{m_Z^4}{16M_2^4} \right]^{3/2}
\left[ 1 - \frac{m_Z^2}{4M_2^2} \right]^{-2},
\end{eqnarray}
}

On the other hand, when the initial state forms a spin-1 state, it annihilates into $W^+ W^-$, $Z h$, and $f \bar{f}$. Below, notations $Q_f$ and $I_{3f}$ denote electric charge, and $SU(2)_L$ charge, respectively, while $k_f$ is defined as $k_f \equiv M_2 [ 1 - m_f^2/M_2^2 ]^{1/2}$.
{\small
\begin{eqnarray}
\sigma_0 v|_{WW} &=&
\frac{\pi \alpha_2^2}{12M_2^2}
\left[ 1 - \frac{m_W^2}{M_2^2} \right]^{3/2}
\left[ 1 - \frac{m_W^2}{2M_2^2} \right]^{-2}
\left[ 1 - \frac{m_Z^2}{4M_2^2} \right]^{-2}
\nonumber \\ && \times
\left[ 1 - \frac{m_Z^2 - m_W^2}{2M_2^2} \right]^{2}
\left[ 1 + \frac{5m_W^2}{M_2^2} + \frac{3m_W^4}{4M_2^4} \right],
\\
\sigma_0 v|_{Zh} &=&
\frac{\pi \alpha_2^2}{12M_2^2}
\left[ 1 - \frac{m_Z^2}{4M_2^2} \right]^{-2}
\left[\left( 1 + \frac{m_Z^2 - m_h^2}{4M_2^2} \right)^2
+ 2 \frac{m_Z^2}{M_2^2} \right]
\nonumber \\ && \times
\left[ 1 - \frac{m_Z^2+m_h^2}{2M_2^2}
+ \frac{\left( m_Z^2 - m_h^2 \right)^2}{16M_2^4} \right]^{1/2},
\\
\sigma_0 v|_{f \bar{f}} &=&
\frac{\pi \alpha_2^2}{6M_2^2}\frac{k_f}{M_2}
\left[4 s_W^4 Q_f^2 \left( 3 - \frac{k_f^2}{M_2^2} \right)
- 3 \frac{m_f^2}{M_2^2} I_{3f}^2
\left( 1 - \frac{m_Z^2}{4M_2^2} \right)^{-2} \right.
\nonumber \\ &&
+ 4 s_W^2 Q_f \left( I_{3f}- 2 s_W^2 Q_f \right)
\left( 3 - \frac{k_f^2}{M_2^2} \right)
\left( 1 - \frac{m_Z^2}{4M_2^2} \right)^{-1}
\nonumber \\ &&
\left. + 4 \left( s_W^4 Q_f^2 - I_{3f} s_W^2 Q + I_{3f}^2/2 \right)
\left( 3 - \frac{k_f^2}{M_2^2} \right)
\left( 1 - \frac{m_Z^2}{4M_2^2} \right)^{-2} \right].
\end{eqnarray}
}

\subsection{Transition between $\widetilde{W}^0 \widetilde{W}^0$ and $\widetilde{W}^+ \widetilde{W}^-$}

Since $\widetilde{W}^0 \widetilde{W}^0$ and the spin-0 state of $\widetilde{W}^+ \widetilde{W}^-$ have the same quantum number, they are mixed each other. In order to evaluate the Sommerfeld factor for the states, we have to calculate the imaginary part of the transition amplitude between $\widetilde{W}^0 \widetilde{W}^0$ and $\widetilde{W}^+ \widetilde{W}^-$. As can be easily understood from the interaction of the neutral wino, the intermediate state of the amplitude is the $W$ boson pair, which is evaluated as
{\small
\begin{eqnarray}
\sigma_0 v|_ {\widetilde{W}^0 \widetilde{W}^0 \leftrightarrow \widetilde{W}^+ \widetilde{W}^- \to WW} =
\frac{2\pi \alpha_2^2}{M_2^2}
\left[ 1 - \frac{m_W^2}{M_2^2} \right]^{3/2}
\left[ 1 - \frac{m_W^2}{2M_2^2} \right]^{-2}.
\end{eqnarray}
}


\end{document}